\documentclass[12pt]{iopart}

  \usepackage{cite}
  \usepackage{graphicx}

\usepackage{iopams}  
\begin{document}

\title[Order-to-chaos transition]{Order-to-chaos transition in the model of quantum pendulum subjected to noisy perturbation}


\author{D.~V.~Makarov, L.~E.~Kon'kov}

\address{
Laboratory of Nonlinear Dynamical Systems,\\
Pacific Oceanological Institute of the Russian Academy of Sciences,\\
43 Baltiiskaya st., 690041 Vladivostok, Russia, URL: http://dynalab.poi.dvo.ru}  

\begin{abstract}

Motion of randomly-driven quantum nonlinear pendulum is considered.
Utilizing one-step Poincar\'e map, we demonstrate that classical phase space
corresponding to a single realization of the random perturbation
can involve domains of finite-time stability.
Statistical analysis of the finite-time evolution operator (FTEO) is carried out
in order to study influence of finite-time stability on quantum dynamics.
It is shown that domains of finite-time stability give rise to ordered
patterns in distributions of FTEO eigenfunctions. 
Transition to global chaos is accompanied by smearing of these patterns;
however, some of their traces survive on relatively long timescales.


\end{abstract}

\vspace{2pc}


\section{Introduction}

It is well-known that deterministic classical systems with few
degrees of freedom can exhibit complicated chaotic behaviour that is 
similar to behaviour of systems with intrinsic random fluctuations.
However, 
the methods used for studying deterministic chaotic systems
and systems under stochastic driving differ substantially.
Representing the external perturbation as some random process,
we add uncertainty into the equations of motion.
Each realization of the perturbation creates an unique
trajectory of the system, therefore, efficient description of the 
system's response implies the usage of statistical analysis.

Nevertheless, single realizations of the random perturbation can be considered
as deterministic functions. As long as the temporal Fourier spectrum of the perturbation
is broad, there is no way for survival of impenetrable stable domains in phase space \cite{WW-TB,Abdullaev11}.
However, statistical analysis of the finite-time Lyapunov exponents in randomly-driven 
systems shows that remarkable fraction of phase space area maintains stability on timescales
which exceed significantly the so-called Lyapunov time
\begin{equation}
 t_\mathrm{L}=\frac{1}{\lambda_L},
\end{equation}
where $\lambda_L$ is the global Lyapunov exponent calculated with $t\to\infty$ \cite{WT01,Laffargue2013}.
Such trajectories form bundles like branched flows in quantum point contacts \cite{Topinka2000,Topinka2001,Kaplan-PRL02,BoLiu13},
or coherent ray clusters in ocean acoustics \cite{Chaos}.
One shouldn't confuse this kind of coherent phenomena with the domains of particle clusterization
in random fields \cite{Klyatskin_Gurarie}.
The main difference is the location of initial conditions for particles.
In the case of the coherent clusters, the initial conditions belong to continuous
manifolds in phase space. 
Location of that manifolds depends on particular realization of the perturbation.
Such manifolds can be found out in various ways. For example, one can calculate
the map of finite-time Lyapunov exponents in phase space \cite{OM,Finn_Apte-Chaos},
or compute eigenfunctions of the transfer operator \cite{Froyland-PRL}.
It turns out that system's behaviour under a single realization can significantly
differ from the picture obtained via statistical averaging.
There can occur phenomena which are typical for deterministic systems, for example,
intermittency and capturing into dynamical traps \cite{Zas,Traps,Tomc_Laksh}.
Thus, one needs some general approach which can be used
for both deterministic and noisy cases.

The problem of interrelation between deterministic and statistical approaches
also arises in quantum systems whose classical counterparts exhibit chaotic behaviour.
Periodic orbit theory \cite{Gutzwiller,Stockman} provides classical interpretation of
quantum spectra and, in addition, reveals some non-classical features,
like scars of wavefunctions \cite{Heller}.
One can suggest that some peculiarities of deterministic quantum systems should 
manifest themselves in quantum systems involving classical noise.
The issue of particular interest is how deterministic phenomena associated with periodic orbits
are revealed under stochastic driving, when there is no periodic orbits in the strict sense.

In the present paper, we demonstrate approach that allows one to analyze quantum systems under weak random perturbation
in the framework of deterministic theory.
In that approach, it is implied that system's behaviour possesses some features which are common for all typical 
realizations of the perturbation.
In this way, any realization is treated as a deterministic process with known spectral properties.
On the classical level, our approach is based on the one-step Poincar\'e map originally introduced in \cite{JPA,PRE73}.
It serves as a tool for finding out phase space patterns repeating themselves after some time interval.
Some of these patterns have regular form and can be referred to as domains of finite-time stability.
We use the quantum counterpart of the one-step Poincar\'e map, the so-called finite-time evolution operator, 
for exploring manifestations of finite-time stability in quantum motion.
Mathematically equivalent approach was used for studying sound propagation in 
a randomly-inhomogeneous oceanic waveguide \cite{UFN,PRE87}. 
In the present paper, we use this approach for a purely quantum problem, namely for the quantum nonlinear pendulum
subjected to broadband perturbation. 
Our main goal is to study how domains of finite-time stability influence quantum dynamics.

It is worth mentioning that the condition of invariance under translation over finite time interval
is too restrictive, and there can be phase space patterns which 
don't satisfy it but correspond to non-chaotic dynamics.
A striking example is the so-called branched flows in quantum point contacts \cite{Kaplan-PRL02,BoLiu13}.
Our approach is rather designed for oscillatory motion subjected to external noise.
In particular, it can be implemented in the problems of noise-driven dissociation or ionization \cite{Singh_Kenfack_Rost-PRA08,Kenfack08-NJP,Feng_Chu}, 
where regular phase space domains prevent transitions into unbounded states, therefore,  
efficient destruction of these domains is of great importance.
The same problem arises in randomly-driven quantum ratchets with cold atoms \cite{PLA}.

The paper is organized as follows. 
The next section describes the model under consideration.
In section \ref{Onestep}, we study classical motion of a randomly-driven pendulum by means of 
one-step Poincar\'e map.
Signatures of classical finite-time stability in quantum dynamics are explored 
in section \ref{FTEO}. In Conclusion, we summarize the results obtained.

\section{Model}
\label{Model}

Consider the quantum Hamiltonian 
\begin{equation}
\eqalign{
 &\hat H = -\frac{\hbar^2}{2}\frac{\partial}{\partial x} + U(x) + \varepsilon V(x,t),\\
 U(x) = -&\cos{x}, \quad V(x,t) = f(t)\sin{x} - f(t+\Delta)\cos{x}},
 \label{qH}
\end{equation}
where $\varepsilon\ll 1$, and $f(t)$ is so-called harmonic noise \cite{HN,Anischenko}.
The corresponding Schr\"odinger equation reads
\begin{equation}
 i\hbar\frac{\partial\Psi}{\partial t} = \hat H\Psi.
 \label{shrod}
\end{equation}
In the present paper we simplify the analysis by considering only quantum states with zero quasimomentum.
It corresponds to the periodic boundary conditions $\Psi(-\pi,t)=\Psi(\pi,t)$. 
This simplification is quite reasonable if we deal with semiclassical regime when energy bands are flat, and tunneling between 
neighbouring potential wells is fairly weak. It demands the Planck constant to be small, therefore,
we set $\hbar=0.1$.
The model (\ref{qH}) was used in \cite{PLA} in the context of quantum ratchet phenomena.

Harmonic noise is described by coupled stochastic differential equations
\begin{equation}
 \dot f=y,\quad
 \dot y=-\Gamma y-\omega_0^2f + \sqrt{2\beta\Gamma}\xi(t),
 \label{ou2d}
\end{equation}
where $\Gamma$ is a positive constant, and
$\xi(t)$ is Gaussian white noise.
The terms $f(t)$ and $f(t+\Delta)$ in (\ref{qH})
correspond to identical realizations of harmonic noise and
differ only by the temporal shift $\Delta$.
The first two moments of harmonic noise are given by 
\begin{equation}
 \left<f\right>=0,\quad
 \left<f^2\right>=\frac{\beta}{\omega_0^2}.
\end{equation}
We set $\beta=1$, that is, the perturbation strength is solely determined 
by the parameter $\varepsilon$.
In the case of low values of $\Gamma$,
the power spectrum of harmonic noise has the peak at the frequency
\begin{equation}
 \omega_{\mathrm{p}}=\sqrt{\omega_0^2-\frac{\Gamma^2}{2}}.
\label{w_p}
\end{equation}
Width of the peak is given by the formula
\begin{equation}
\Delta\omega = \sqrt{\omega_{\mathrm{p}}+\Gamma\omega'}-
\sqrt{\omega_{\mathrm{p}}-\Gamma\omega'},
 \label{width}
\end{equation}
where $\omega'=\sqrt{\omega_0^2-\Gamma^2/4}$.
As $\Gamma\to 0$, $f(t)\to \sin(\omega_0t+\phi_0)$, where $\phi_0$ is determined
by initial conditions in (\ref{ou2d}). 
Setting $f(0)=1$, $y(0)=0$, and 
$ \Delta=\pi/(2\omega_0)$,
one can easily find that
$V(x,t)=\sin(x+\omega_0t)$ in the case of $\Gamma=0$. 
Hence, it turns out that, for $\Gamma>0$, $V(x,t)$
behaves like a plane wave whose amplitude and 
phase velocity fluctuate with time. The plane-wave acts as a dragging 
force for particles and leads to the onset of directed current, i.~e. the ratchet effect.
This kind of ratchets is known as travelling potential ratchets \cite{JETPL,PRE75,JTPL08,EPJB}.
In the semiclassical regime, direction of the current coincides with the direction of 
the perturbation phase velocity, provided dynamical barriers preventing the transition of particles 
into infinite regime are destroyed. More intricate behaviour is observed in the deep quantum regime \cite{Fromhold-PRA13}.
In the present work we use the following values
of parameters: $\omega_0=1$, $\Gamma=0.1$ and $\varepsilon=0.05$.

\section{One-step Poincar\'e map}
\label{Onestep}

Let's begin with the classical level and consider the classical counterpart of the Hamiltonian (\ref{qH})
\begin{equation}
H=\frac{p^2}{2}+U(x)+\varepsilon V(x,\,t).
\label{ham-st}
\end{equation}
Corresponding equations of motion read
\begin{equation}
\frac{dx}{dt}=p,\quad
\frac{dp}{dt}=-\frac{dU}{dx}-\varepsilon\frac{dV}{dx}.
\label{sys-xp}
\end{equation}
Let's consider some arbitrarily chosen realization of $V(x,t)$.
Then we can treat $V(x,t)$ as a deterministic function and
refer to (\ref{sys-xp}) as the system of ordinary differential equations.
As $V(x,t)$ is an oscillating function of time, the domains of finite-time stability
may involve components which transform to themselves without mixing in course of evolution from $t=0$
to $t=\tau$.
These components can be found out by means of the one-step Poincar\'e map \cite{JPA,PRE73,Gan1}
\begin{equation}
p_{i+1}=p(t=\tau;\,p_i,x_i),\quad
x_{i+1}=x(t=\tau;\,p_i,x_i),
\label{map}
\end{equation}
where $p(t=\tau;\,p_i,x_i)$ and $x(t=\tau;\,p_i,x_i)$ are
solutions of  (\ref{sys-xp}) with initial conditions
 $p(t=0)=p_i$, $x(t=0)=x_i$. 
 \begin{figure}[!htb]
\begin{center}
\includegraphics[width=0.3\textwidth]{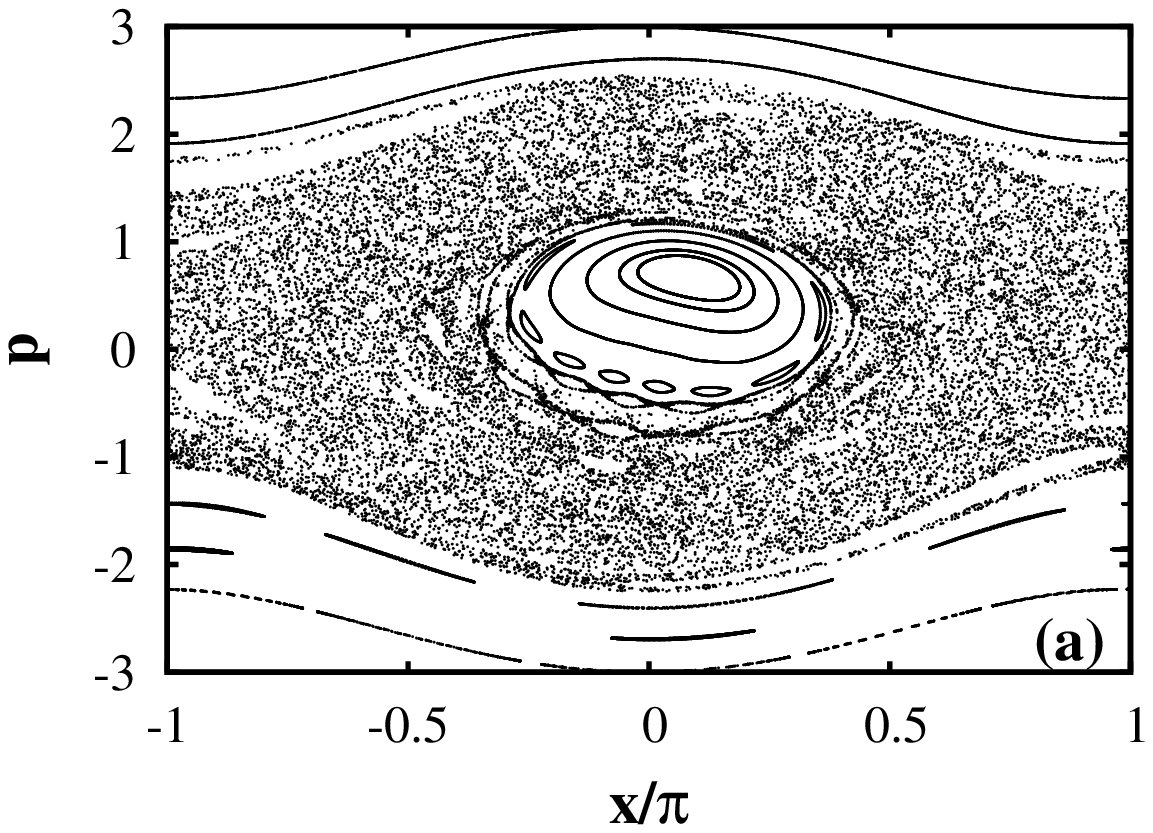}
\includegraphics[width=0.3\textwidth]{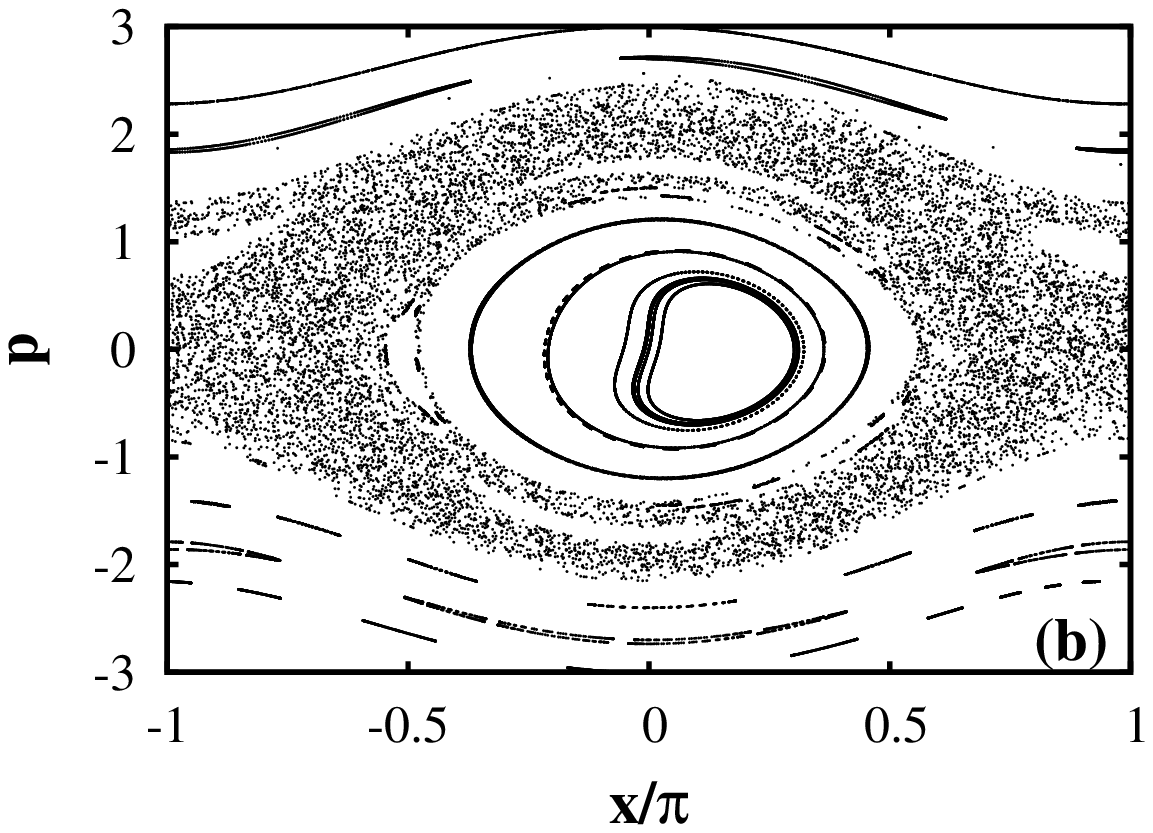}\\
\includegraphics[width=0.3\textwidth]{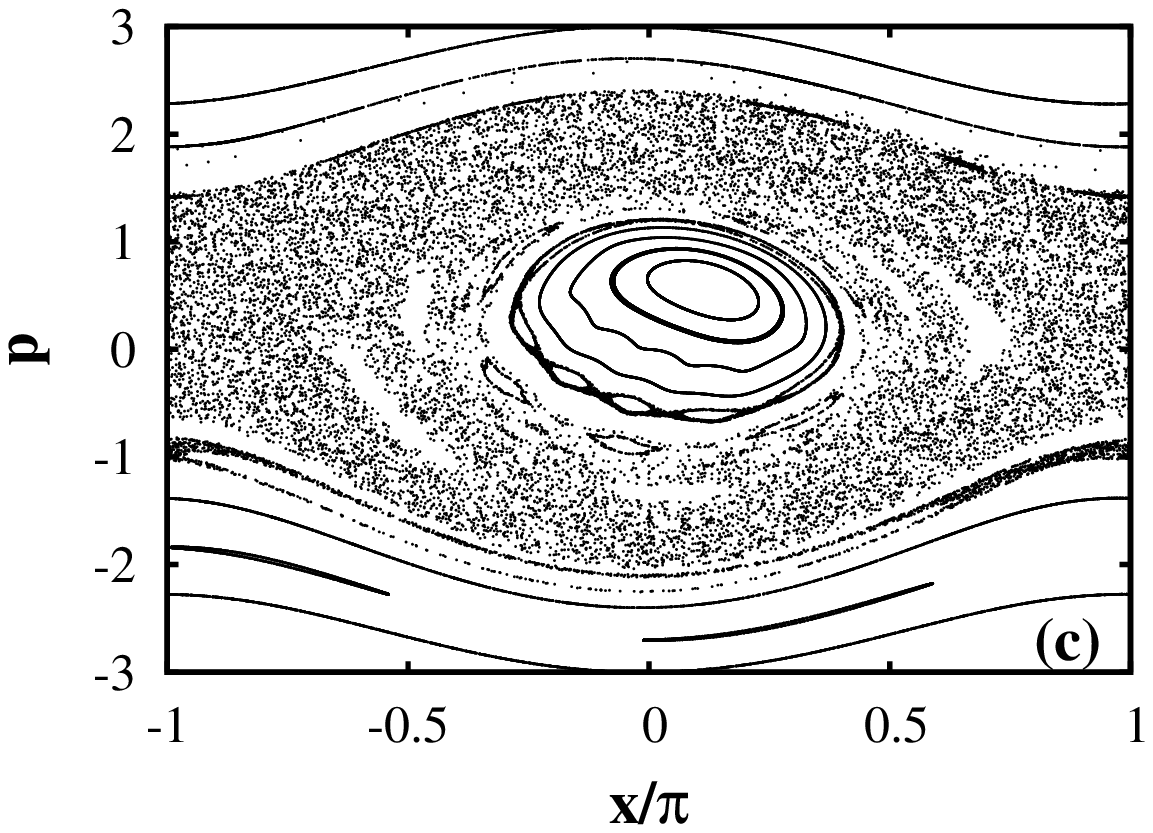}
\includegraphics[width=0.3\textwidth]{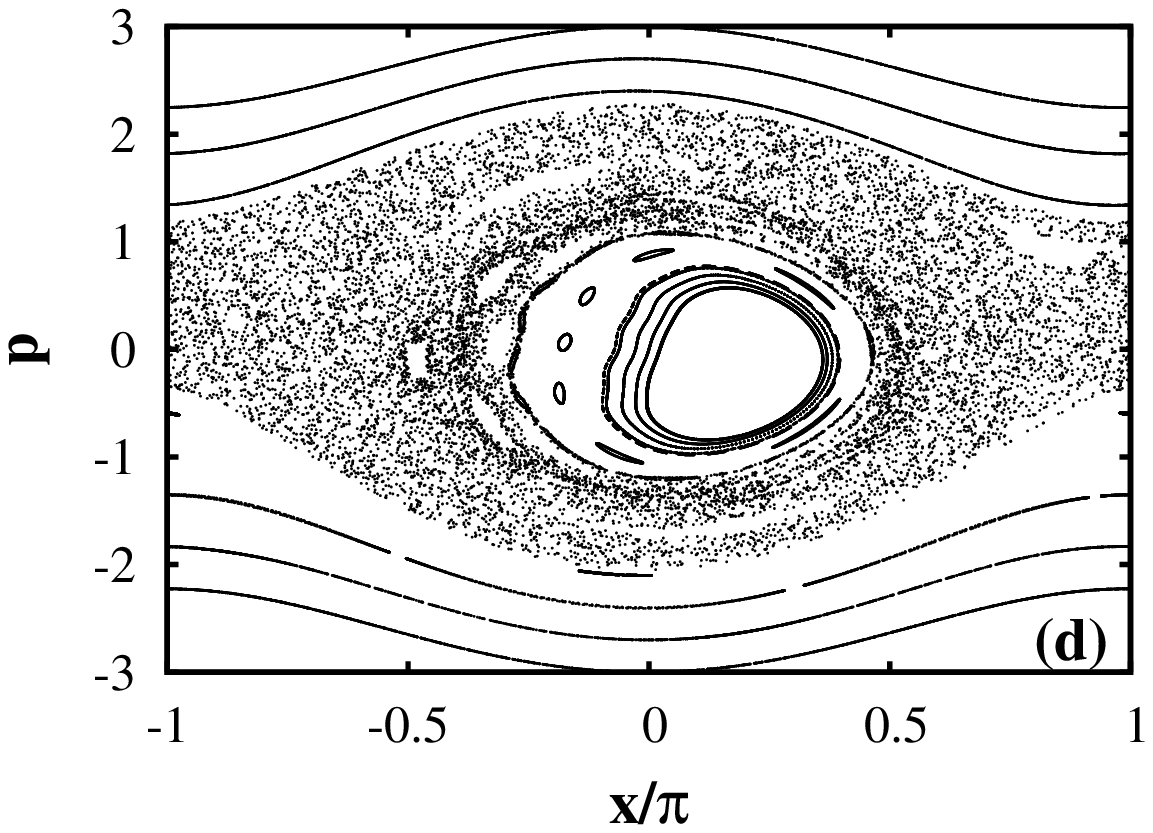}
\caption{Phase space portraits constructed via one-step Poincar\'e map with $\tau=4\pi$.
Figures (a)-(d) correspond to different realizations of harmonic noise.
}
\label{fig-p04}
\end{center}
\end{figure}
 One-step Poincar\'e map is equivalent to the usual Poincar\'e map
with the Hamiltonian
\begin{equation}
\bar H=\frac{p^2}{2}+U(x)+\varepsilon\tilde V(x,\,t),
\label{ham-period}
\end{equation}
\begin{equation}
\tilde V(x,\,\bar t+n\tau)=V(x,\,\bar t),\quad
0\leqslant \bar t\leqslant\tau,
\label{xi-sr}
\end{equation}
$n$ is an integer. 
Thus, we replace the original system by the equivalent 
time-periodic one. Validity of this replacement is provided by the time restriction 
to the interval $[0:\tau]$. An alternative approach for generalization of Poincar\'e map
onto stochastic dynamical systems was offered in \cite{Hitczenko,Berglund}.

\begin{figure}[!htb]
\begin{center}
\includegraphics[width=0.3\textwidth]{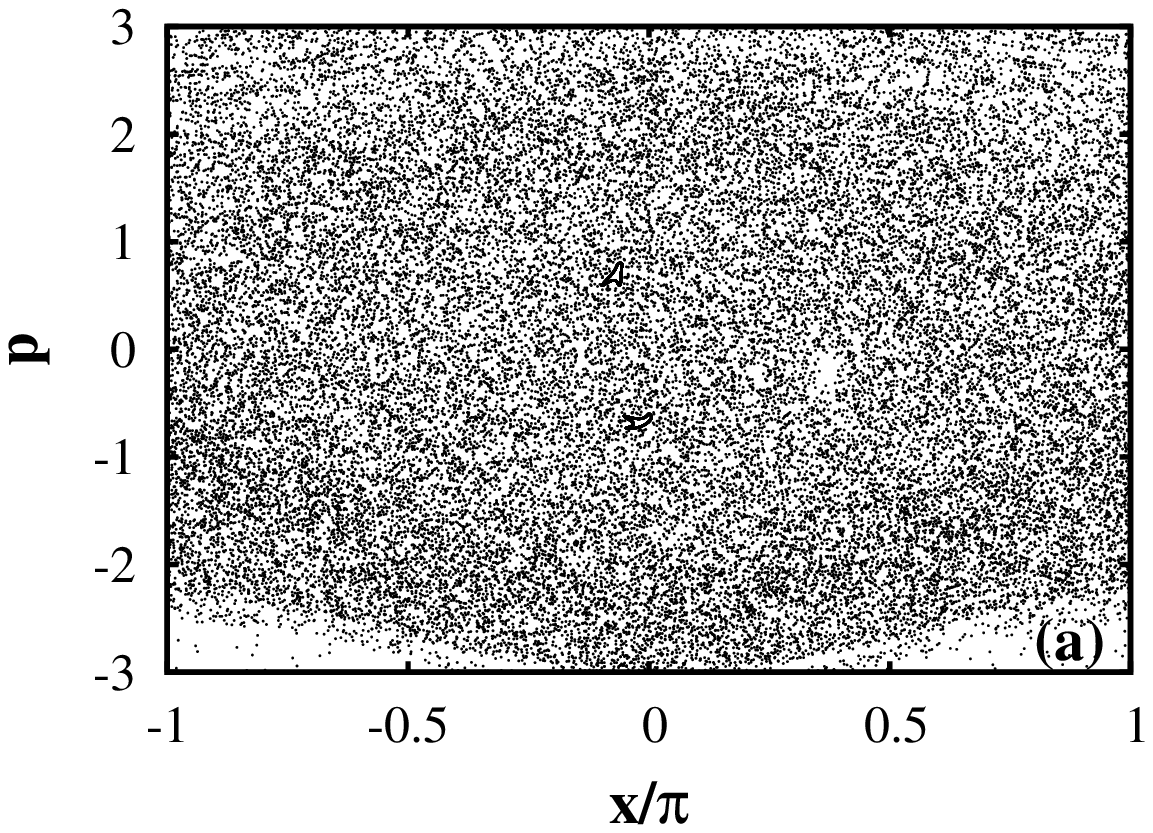}
\includegraphics[width=0.3\textwidth]{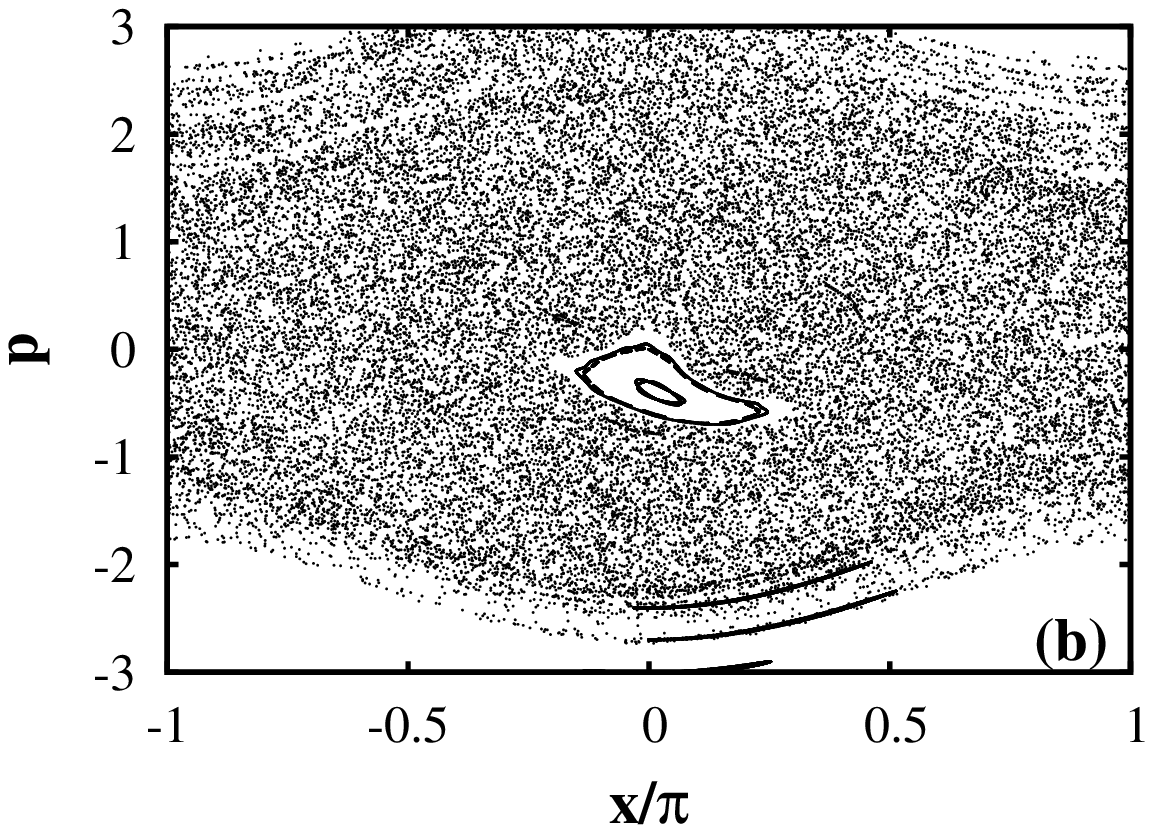}\\
\includegraphics[width=0.3\textwidth]{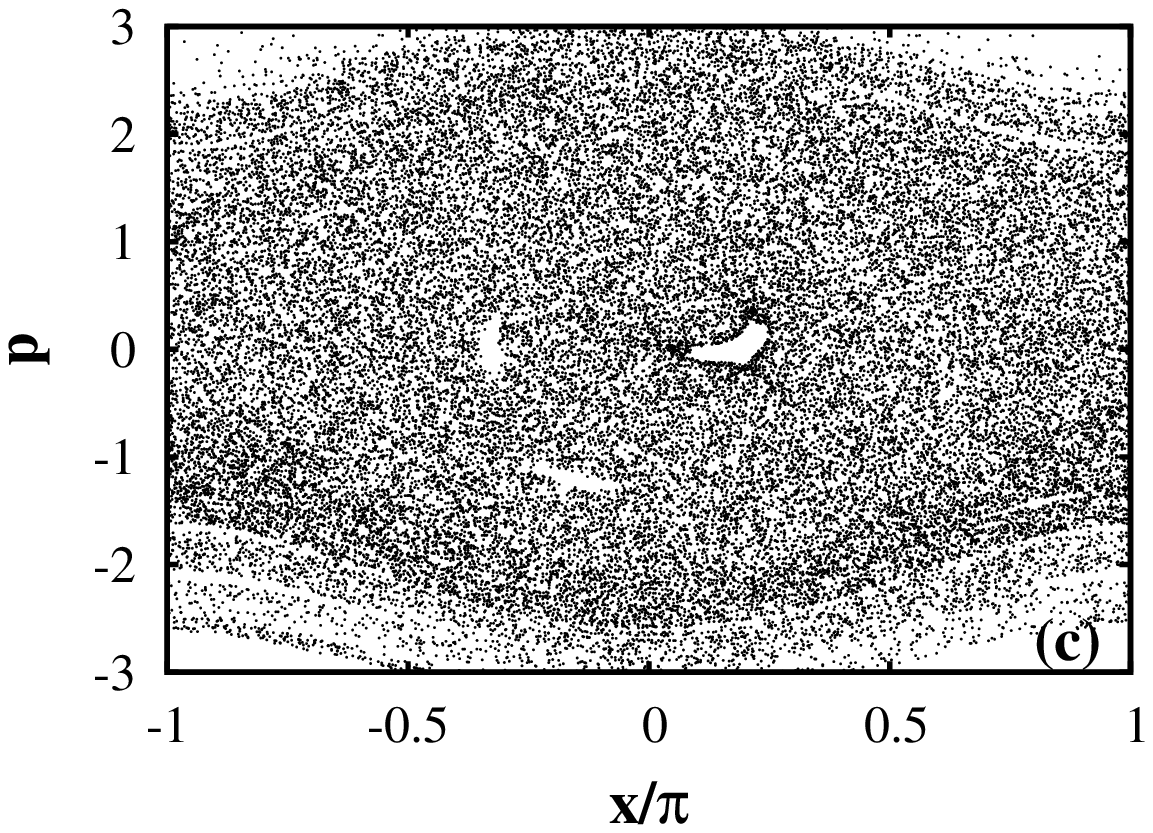}
\includegraphics[width=0.3\textwidth]{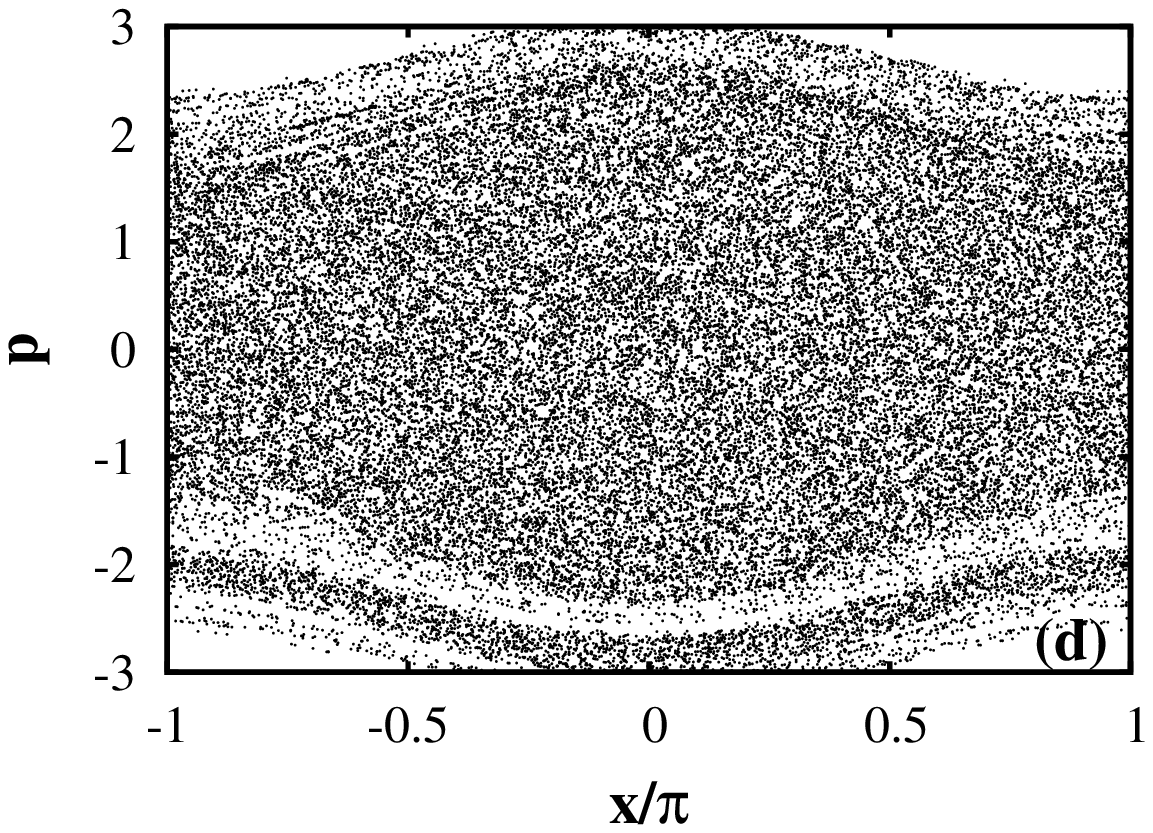}
\caption{Phase space portraits constructed via one-step Poincar\'e map with $\tau=20\pi$.
Figures (a)-(d) correspond to different realizations of harmonic noise.
}
\label{fig-p20}
\end{center}
\end{figure}

Following the analogy with the usual Poincar\'e map, one can deduce the main property of one-step Poincar\'e map:
{\em if a trajectory of (\ref{map}) forms a closed continuous curve in phase space,
any point belonging to it is an initial condition for a trajectory of (\ref{sys-xp}) that
remains stable at $t=\tau$.}
The inverse statement is not generally true, therefore, one-step Poincar\'e map 
yields a sufficient but not necessary criterion of stability.
This means that one-step Poincar\'e map basically underestimates area of regular domains.

As we artificially reduce the problem to a time-periodic one,
theory of time-periodic Hamiltonian systems can be invoked \cite{JPA,PRE73}.
Phase space structure of one-step Poincar\'e map is determined
by resonances 
\begin{equation}
 m_1\tau = m_2T,
 \label{rescond}
\end{equation}
where $T$ is the period of unperturbed motion. Period is function of the action variable  defined as \cite{Zas} 
\begin{equation}
 I = \frac{1}{2\pi}\oint p\,dx.
\end{equation}
Each resonance has certain width in the space of the action.
Width can be calculated using the theory of nonlinear resonance. It can be shown that the distance between neighbouring dominant nonlinear
resonances decreases with increasing $\tau$ as $\tau^{-1}$ \cite{JPA}. This results in resonance overlapping
and gradual transition to global chaos according to the Chirikov criterion \cite{Chirikov79}.

Figures \ref{fig-p04} and \ref{fig-p20} demonstrate phase portraits constructed by means of map (\ref{map}) 
with $\tau=4\pi$ and $\tau=20\pi$, respectively. Notably, figures corresponding to the same value of $\tau$
but different realizations of harmonic noise represent very similar patterns with nearly the same fraction of regular phase space area.
In the case of $\tau=4\pi$ (see Fig.~\ref{fig-p04}), the central part of phase space 
maintains stability for all realizations of harmonic noise.
This domain corresponds to the vicinity of the stable equilibrium point of the unperturbed system.
Chaos emerges in the neighbourhood of the unperturbed separatrix.
Increasing of $\tau$ to $20\pi$ results in remarkable shrinking of the regular phase space area,
as it is illustrated in Fig.~\ref{fig-p20}. 
Phase space region corresponding to finite motion becomes submerged into the 
chaotic sea. Stable domains inside the unperturbed separatrix may survive only as small islands.
For some realizations, they disappear completely. For example, the internal part of the chaotic sea presented
in Fig.~\ref{fig-p20}(d) is almost uniform, without any apparent islands.
Further increasing of  $\tau$ results in complete disappearance of stable islands in
the phase space region enclosed by the unperturbed separatrix.

\section{Finite-time evolution operator}
\label{FTEO}

Phase space portraits presented in the preceding section indicate fast destruction of stable domains.
Let's consider how this process is revealed in quantum dynamics.
Quantum counterpart of one-step Poincar\'e map is the  operator $\hat G$ defined as
\begin{equation}
 \hat G(\tau)\bar\Psi(x)=\exp\left(-\frac{i}{\hbar}\hat{H}\tau\right)\bar\Psi(x)
 =\left.\Psi(x,\,t)\right\vert_{t=\tau},
\label{evolution}
\end{equation}
where $\bar{\Psi}(x)=\Psi(x,\,t=0)$.
Hereafter we shall refer to  $\hat G$ as the finite-time evolution operator (FTEO).
Each realization of harmonic noise creates its own realization of the FTEO.
FTEO was firstly utilized in \cite{Kol97} for the problem of noise-driven quantum diffusion.
Wave analogue of the FTEO was used in  \cite{Levelspacing,Tomc11,UFN,PRE87,Hege13}.

Peculiarities of classical phase space should be reflected in spectral properties
of the FTEO.
Eigenvalues and eigenfunctions of the FTEO satisfy the equation
\begin{equation}
\hat G\Psi_m(x)=g_m\Psi_m(x)\equiv e^{-i\epsilon_m/\hbar}\Psi_m(x).
\label{eigen}
\end{equation}
Quantity $\epsilon_m$ is the analogue of quasienergy in time-periodic quantum systems.
As increasing of $\tau$ results in destruction of regular domains, one should expect
transition in statistics of level spacings $s=\epsilon_{m+1}-\epsilon_m$ from
Poissonian to Wigner-like regime \cite{Kol97}. This expectation can fail in the presence 
of periodic-orbit bifurcations \cite{Levelspacing,UFN}.
Moreover, analysis of level spacing statistics doesn't provide
accurate estimate of the regular phase space area \cite{PRE87}.
In this way, analysis of FTEO eigenfunctions seems to be more robust way.
To facilitate the analysis,
eigenfunctions $\Phi_m$  can be expanded over eigenstates of the unperturbed potential
\begin{equation}
 \Phi_m(x) = \sum\limits_n c_{mn}\phi_n(x).
 \label{fl_expand}
\end{equation}
Chaos implies extensive  transitions between energy levels \cite{BerKol}, therefore,
a chaos-assisted eigenfunction of FTEO should be compound of many unperturbed eigenstates.
Thus, one can use the participation ratio 
\begin{equation}
\nu = \left(
\sum\limits_{m}\vert c_{mn}\vert^4
\right)^{-1},
 \label{npc}
\end{equation}
as  measure of ``chaoticity''. 
Phase space region associated with an eigenfunction can be found by means of the parameter \cite{Viro05}
\begin{equation}
\mu=\sum\limits_{m=1}^M\vert c_{mn}\vert^2m.
 \label{muu}
\end{equation}
Indeed, the formula $\left<I\right>=\hbar(\mu+1/2)$ yields the expectation value of the classical action corresponding to the eigenfunction. 
Parameters $\nu$ and $\mu$ provide suitable classification of eigenfunctions and can be used for tracking the transition from order
to chaos by means of numerical simulation.

\begin{figure}[!htb]
\begin{center}
\includegraphics[width=0.34\textwidth]{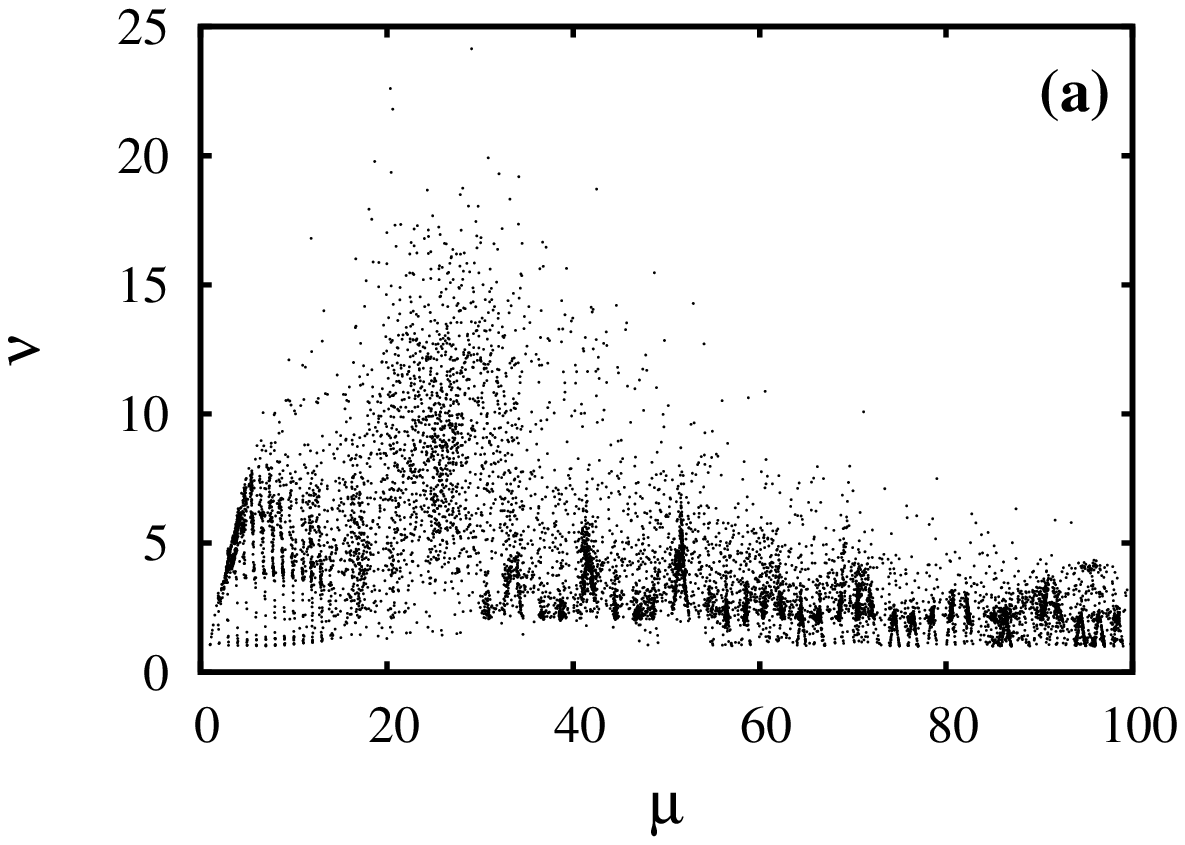}
\includegraphics[width=0.34\textwidth]{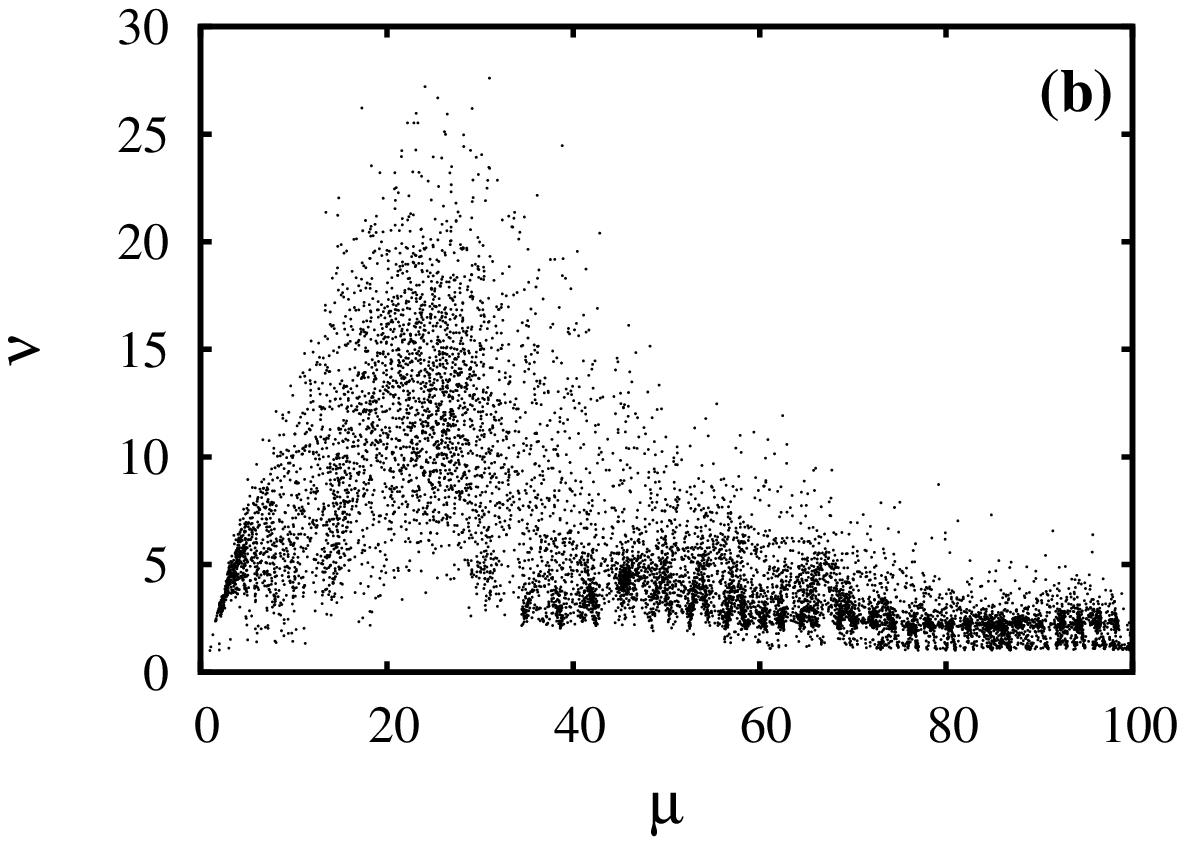}\\
\includegraphics[width=0.34\textwidth]{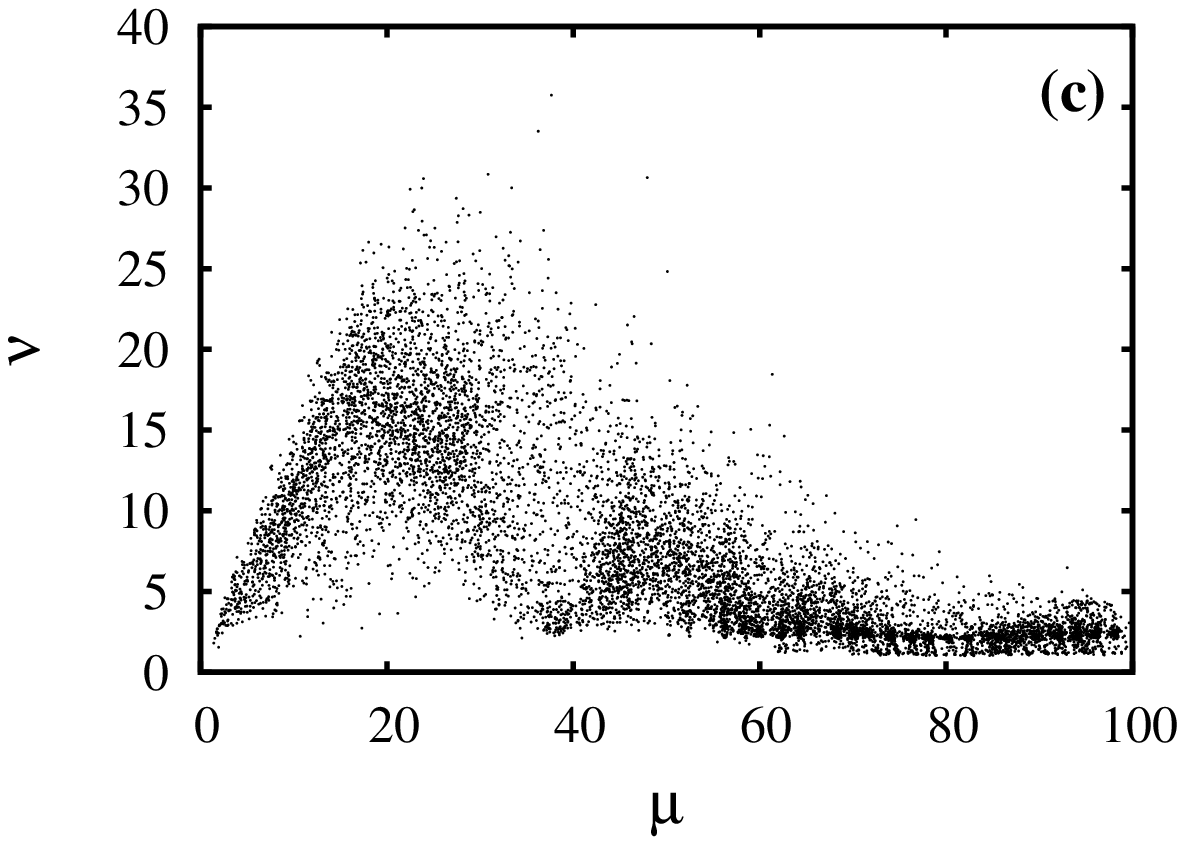}
\includegraphics[width=0.34\textwidth]{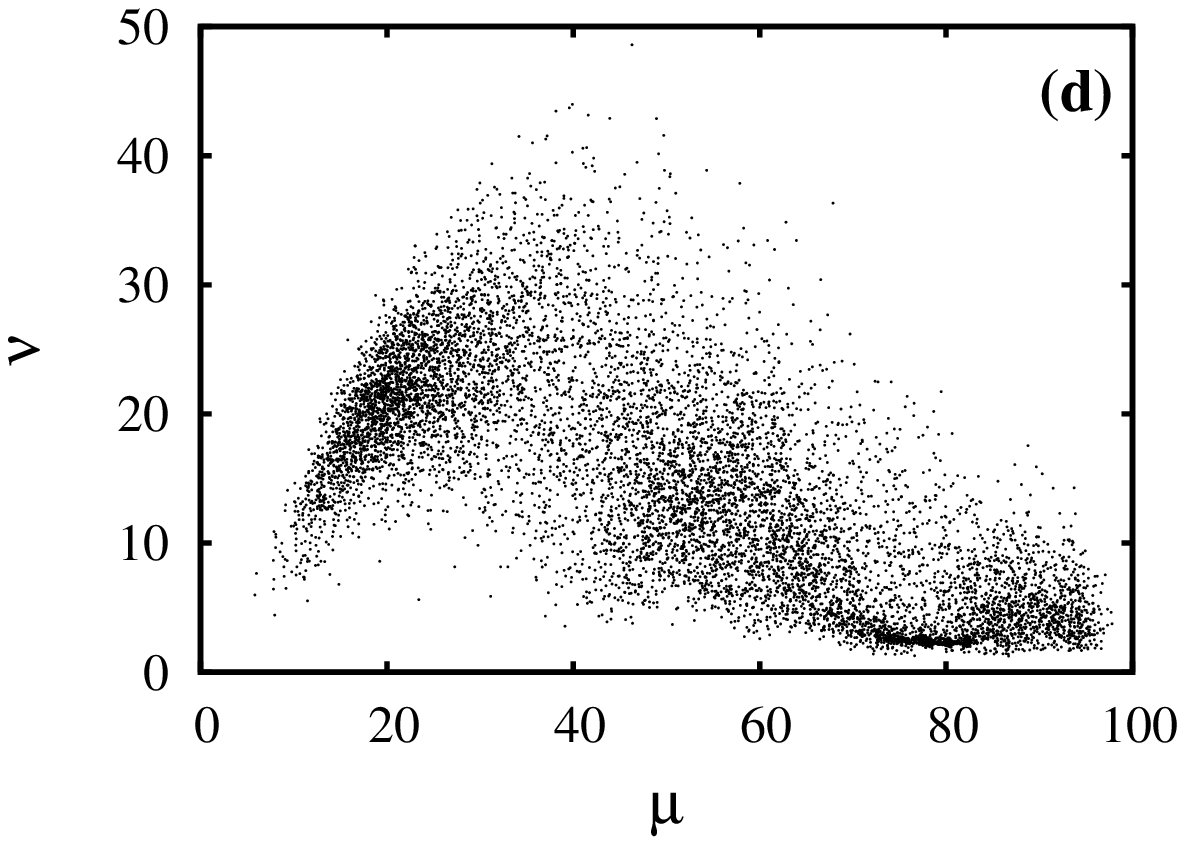}
\caption{Distribution of FTEO eigenfunctions in the  $\mu$-$\nu$ plane.
Values of $\tau$:
(a) $4\pi$, (b) $10\pi$, (c) $20\pi$, (d) $100\pi$.
}
\label{fig-munu}
\end{center}
\end{figure}
%


Numerical simulation was conducted with the ensemble of 100 realizations of the FTEO.
100 eigenfunctions with the lowest values of $\mu$ are taken into account
for each realization.
An informative view is provided by distributions of FTEO eigenfunctions 
in the $\mu$-$\nu$ space.
These distributions corresponding to different values of $\tau$ 
are presented in Fig.~\ref{fig-munu}. 
For relatively small values of $\tau$, the dots corresponding to eigenfunctions form ordered patterns
consisted of distinct slightly biased lines. Such patterns were earlier observed 
in \cite{PRE87}, where they were called ``stalagmites''. Each ``stalagmite'' is formed by eigenfunctions
localized near periodic orbits of the one-step Poincar\'e with the same location in the action space.
Transition to chaos is accompanied by delocalization of eigenfunctions and smearing of 
``stalagmites''. It should be noted that smearing is partially suppressed by dynamical localization \cite{Stockman},
that is, weakly unstable periodic orbits retain the ability to trap eigenfunctions \cite{PRE76}.
For $\tau=4\pi$, stalagmites occur in the whole range of $\mu$ values, except for the vicinity of $\mu=25$.
This value of $\mu$ corresponds to the phase space region near the unperturbed separatrix of the pendulum,
where the classical chaotic sea is originated initially.
As $\tau$ increases, the smeared domain grows indicating gradual transition to chaos due to overlapping
of classical resonances (\ref{rescond}).
Notably, chaos-assisted delocalization firstly emerges in the range of small values of $\mu$, corresponding to finite motion. 
This infers efficient destruction of invariant curves impeding transition between finite
and infinite regimes. Results of Ref.~\cite{PLA} show that the destruction of invariant curves leads to the onset of directed current.

Nevertheless, traces of eigenfunctions with good persistence to chaos are visible even for $\tau=100\pi$.
For instance, there is a small slightly smeared horizontal stripe near $\mu \simeq 80$ (see Fig.~\ref{fig-munu}(d)) 
corresponding to infinite motion with relatively high velocities.
That phase space region is characterized by inequality $T\ll\tau$ that anticipates
weak influence of resonances (\ref{rescond}), and, hence, weakness of chaos induced by their overlapping.

\begin{figure}[!htb]
\begin{center}
\includegraphics[width=0.4\textwidth]{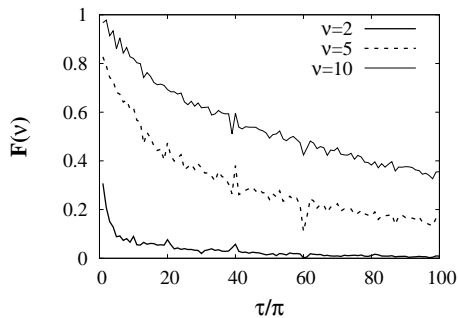}
\caption{
Fractions of eigenfunction ensemble, corresponding to different regimes of localization, vs time.
}
\label{fig-npc}
\end{center}
\end{figure}

As domains of finite-time stability in phase space give rise to FTEO eigenfunctions
with small $\nu$, one can estimate their contribution using the cumulative distribution
\begin{equation}
F(\nu)=\int\limits_{1}^{\nu}\rho(\nu')\,d\nu',
\label{cumul1} 
\end{equation}
where $\rho(\nu')$ is the corresponding probability density function.
We refer to the case $\nu\le 2$ as the regime of strong localization.
Indeed, inequality $\nu\le 2$ implies that a FTEO eigenfunction is mainly contributed from 
one or two unperturbed eigenstates.
In addition, let's consider two regimes of moderate localization: $\nu\le 5$ and $\nu\le 10$.
The sense of the latter two criteria becomes apparent if we take into account that
finite motion corresponds to the 25 lowest unperturbed eigenstates.
Thus, the inequality $\nu\le 5$ ($\nu\le 10$) picks out FTEO eigenfunctions 
occupying less than 20 percents (40 percents) of the phase space area enclosed by the unperturbed separatrix.
As it follows from Fig.~\ref{fig-npc}, fraction of strongly localized eigenfunctions rapidly decays
down to nearly zero. It indicates the absence of significant long-living regular domains in classical phase space.
It should be mentioned that the opposite situation was observed in Ref.~\cite{PRE87}, where slow decay of $F(2)$
was linked to the presence of degenerate tori in phase space. In the present case there is no degenerate tori,
therefore, there is no such route to the persistence of regular domains.
Fractions of moderately localized functions decay much slower, and their impact remains significant
even at $\tau=100\pi$. It can be understood as manifestation of dynamical localization, i.~e.
partial suppression of chaos-assisted diffusion due to the interference.

\section{Conclusion}

In the present paper, we demonstrate approach designed for studying quantum manifestations
of classical finite-time stability under weak external random perturbation.
The approach is based on construction of the finite-time evolution operator (FTEO). 
It is emphasized that statistical properties of 
FTEO eigenfunctions are closely linked to classical phase space structure revealed by one-step
Poincar\'e map that can be regarded as the classical counterpart of the FTEO.
In particular, the stalagmite-like patterns on distributions of eigenfunctions in the space of parameters
$\mu$ and $\nu$ are related to wavefunction concentrations near the periodic orbits of the one-step Poincar\'e map.
Increasing of time results in emergence of chaos, therefore,
periodic orbits loss their stability. This leads to smearing of the  ``stalagmites''. 
Chaos-assisted destruction of ``stalagmites'' represents an alternative view onto the order-to-chaos transition
in randomly-driven quantum systems.

In the present work, we consider stable domains satisfying the condition of invariance
under translation over finite time interval. 
They occur owing to the same mechanism as stable islands in time-periodic systems. 
Namely, they are remnants of nonlinear resonances of one-step Poincare map and arise in somewhat regular and predictable way. 
To underline the latter circumstance, we present phase space portraits corresponding to different realizations, 
and one can see that regular domains appear in the same ranges of the action (or energy) values.

This work is supported by the Russian Foundation of Basic Research under project 13-01-12404, 
and by the Siberian and Far-Eastern Branches of the Russian Academy of Sciences under joint project 12-II-SO-07-022. 
Authors are grateful to Michael Uleysky for the assistance in preparation of the manuscript.

\section*{References}


\providecommand{\newblock}{}

\end{document}